\documentclass[runningheads]{llncs}
\usepackage[T1]{fontenc}
\usepackage{graphicx}
\usepackage{multirow}
\usepackage{longtable}
\usepackage{tcolorbox}
\usepackage{caption}
\usepackage{subcaption}
\begin{document}
\title{Digital Transformation Chatbot (DTchatbot): Integrating Large Language Model-based Chatbot in Acquiring Digital Transformation Needs}
\titlerunning{Digital Transformation Chatbot (DTchatbot)}
%
\author{Jiawei Zheng\orcidID{0000-0002-6515-6423} \and
Gokcen Yilmaz\orcidID{0000-0002-4069-409X} \and \\
Ji Han\orcidID{0000-0003-3240-4942} \and
Saeema Ahmed-Kristensen\orcidID{0000-0003-4035-7939}}
\authorrunning{J. Zheng  et al.}
%
\institute{DigitLab, University of Exeter, UK \\
\email{\{J.Zheng2, G.Yilmaz, J.Han2, S.Ahmed-Kristensen\}@exeter.ac.uk}\\
}
\maketitle              
\begin{abstract}
    Many organisations pursue digital transformation to enhance operational efficiency, reduce manual efforts, and optimise processes by automation and digital tools. To achieve this, a comprehensive understanding of their unique needs is required. However, traditional methods, such as expert interviews, while effective, face several challenges, including scheduling conflicts, resource constraints, inconsistency, etc. To tackle these issues, we investigate the use of a Large Language Model (LLM)-powered chatbot to acquire organisations' digital transformation needs. Specifically, the chatbot integrates workflow-based instruction with LLM's planning and reasoning capabilities, enabling it to function as a virtual expert and conduct interviews. We detail the chatbot's features and its implementation. Our preliminary evaluation indicates that the chatbot performs as designed, effectively following predefined workflows and supporting user interactions with areas for improvement. We conclude by discussing the implications of employing chatbots to elicit user information, emphasizing their potential and limitations.

\end{abstract}

\section{Introduction}

Digital transformation is a strategic imperative for organisations seeking to  integrate digital technologies into different aspects of their business models, which can fundamentally change how organisations operate and deliver value to customers. It is a process that aims to improve an organisation by triggering changes to its properties through combination of information, computing and communication technologies~\cite{DTconcept2019}.  
For example, in inventory management, deploying Internet of Things (IoT) devices such as sensors enables real-time monitoring of stock levels, allowing automated reordering and seamless communication with suppliers. This reduces manual efforts, enhances inventory accuracy, and minimises delays in replenishment~\cite{mastosIndustry40Sustainable2020}.
To ensure effective and efficient digital transformation, we must begin with a comprehensive understanding of an organisation's needs, challenges, and current operational practices.

Currently, organisations' digital transformation needs are typically elicited through one of three methods: 1) workshops or interviews, where information is captured via audio recording devices or handwritten notes; 2) engagement with external digital consultants to provide strategic advice and recommendations; and 3) internal assessments led by organisational staff~\cite{bellantuonoDigitalTransformationModels2021}. While these methods can be effective under certain cases, they also come with several challenges that make them less suitable for the fast-paced and resource-constrained environment of current businesses~\cite{schneiderDigitalTransformationWhat2021,furjanUnderstandingDigitalTransformation2020,cozmiucConsultantsToolsManage2021}. For instance: 

\begin{itemize}
    \item \textbf{Time Consuming}: Data collected during these sessions is typically recorded using voice recording devices or documented as written notes. Transcribing audio recordings into written text or analysing handwritten notes requires significant effort and time, introducing complexity, increasing workloads, and causing delays.
    \item \textbf{High Costs}: Engaging with digital transformation experts can be resource-intensive for organisations, especially for Small and Medium-sized Enterprises (SMEs).
    \item \textbf{Limited Scalability}: Traditional workshops and interviews are inherently one-off events, requiring significant coordination and effort to schedule and execute.
    \item \textbf{Fragmented Communication}: In multilingual organizations, the lack of a shared language or communication platform can hinder meaningful dialogue, leaving critical needs misunderstood or overlooked entirely. Moreover, once workshops or consultations conclude, businesses often lack mechanisms for iterative follow-up or ongoing support. This can impede the organization’s ability to adapt strategies as new challenges emerge.
    
\end{itemize}

In response, online questionnaires have been adopted to elicit information~\cite{xiaoTellMeYourself2020}. While these tools can be provided 24/7 to access and complete in their own space, regardless of geographic location, these questionnaires lack interactive features, which results in survey-taking fatigue~\cite{ben2008respondent}. This problem is exacerbated with open-ended questions, which is an important method to collect valuable information and deeper insights. Formulating and typing responses to such questions requires significant time and effort from respondents, increasing their cognitive burden. As a result, respondents are more likely to skip such questions or provide low-quality or even irrelevant answers. This, in turn, compromises the quality and reliability of the data collected. Additionally, analysing and consolidating responses from open-ended questions is a time-intensive task for researchers.

The advent of chatbots, enhanced by increasingly powerful conversational capabilities, offers a compelling alternative to traditional online questionnaires~\cite{xiaoTellMeYourself2020}. While retaining the benefits of online questionnaires, chatbots introduce an interactive and dynamic dimension that can enhance participant engagement and improve the quality of responses. Recent advancements in Large Language Models (LLMs) further amplify the potential of chatbots, equipping them with diverse domain knowledge and the ability to deliver human-like interactions and effectively manage complex conversations. For instance, LLM-powered chatbots can assist participants by clarifying questions, guiding them through the conversation, and encouraging more thoughtful responses. 

This paper investigates whether LLM-powered chatbots can enhance the user experience in understanding and navigating the digtal transformation of organisations. 
We introduce DTchatbot, a conversational agent designed to guide organizational representatives through a structured dialogue aimed at uncovering their unique challenges, goals, and transformation priorities.
We present the design and development of DTchatbot. The chatbot employs a workflow-based approach to guide users through a series of questions, meticulously designed to identify the challenges, needs, and goals unique to their organizations. To enhance accessibility and usability, the system integrates a Speech-to-Text model that seamlessly transcribes participants' audio responses into text, enabling natural and intuitive interactions. To the best of our knowledge, we are the first to develop and use the LLM-powered chatbot for conducting expert consulting sessions. A preliminary user study with two SMEs and two experts indicates that the DTchatbot effectively conducts conversational interviews and supports insightful user engagement.

\section{Related Work}

Digital transformation is a critical process for modern organisations, involving the adoption of digital technologies to enhance operational efficiency, drive innovation, and improve customer experiences. It brings new business and operating models across all sectors and requires organisations to fundamentally rethink their workflows, strategies, and current practices. Many studies and industries highlight and demonstrate the benefits of digital transformation~\cite{bellantuonoDigitalTransformationModels2021a,schneiderDigitalTransformationWhat2021,furjanUnderstandingDigitalTransformation2020}. For example, studies have shown that adopting IoT technologies can lead to increased operational efficiency, such as reducing production costs through automation or optimising supply chain management~\cite{Taj-iot-basedSupplyChain2023}. The success of digital transformation depends heavily on a deep understanding of an organisation's needs.

Recent advancements in conversational AI and chatbots have paved the way for innovative approaches to understanding users' needs. 
Conversational AI and chatbots have long been of interest to the Human-Computer Interaction (HCI) community due to their various interaction benefits, such as the application of learning and driving assistant ~\cite{caiAdvancingKnowledgeTogether2024,huangChatbotFatiguedDriver2024} and customer service agent~\cite{Xuchatbotcustomeragent2017}. One key advantage is that conversational interfaces provide a natural and intuitive way for users to express themselves, which in turn improves the usability of a system. Unlike traditional online questionnaires and surveys, conversational interactions offer flexibility and can accommodate diverse user requests without forcing users to adhere to predefined workflow~\cite{traum2017computational}. This adaptability makes conversational AI particularly effective for engaging with users and eliciting complex information.

Many researchers have investigated the use of conversational AI for eliciting user information through a one-on-one, text-based conversation. For example, Cranshaw et al.\ introduce a conversational agent designed to assist with scheduling events by eliciting key information from dialogue~\cite{schedulingagentJustin2017}. Moreover, conversational AI has been successfully applied in job interview settings~\cite{Liinterviewjobcandidates2017}. 
These findings highlighted the potential of conversational AI to perform human-liked interactions and collect nuanced and contextually rich data. 

Building on existing work, we focus on investigating the use of a chatbot to conduct interviews with organisational stakeholders, Our aim is to facilitate the elicitation of information crucial for guiding their digital transformation. 

\section{DTchatbot features}

Instead of building the chatbot into existing chat applications, such as WhatsApp, Slack, etc., and other third-party chatbot platforms, e.g., juji~\cite{xiaoTellMeYourself2020}, we decided to implement a fully web-based platform. This was guided by: 1) it has the freedom to implement new functionalities and allows for better adaptability, since existing platforms often come with inherent limitations, including restrictions on customisations, data access, and integration with external tools, and 2) these platforms are often closed source, thus leaving little opportunity to implement and test our own functionalities. 

\begin{figure}
    \centering
    \includegraphics[width=0.8\linewidth]{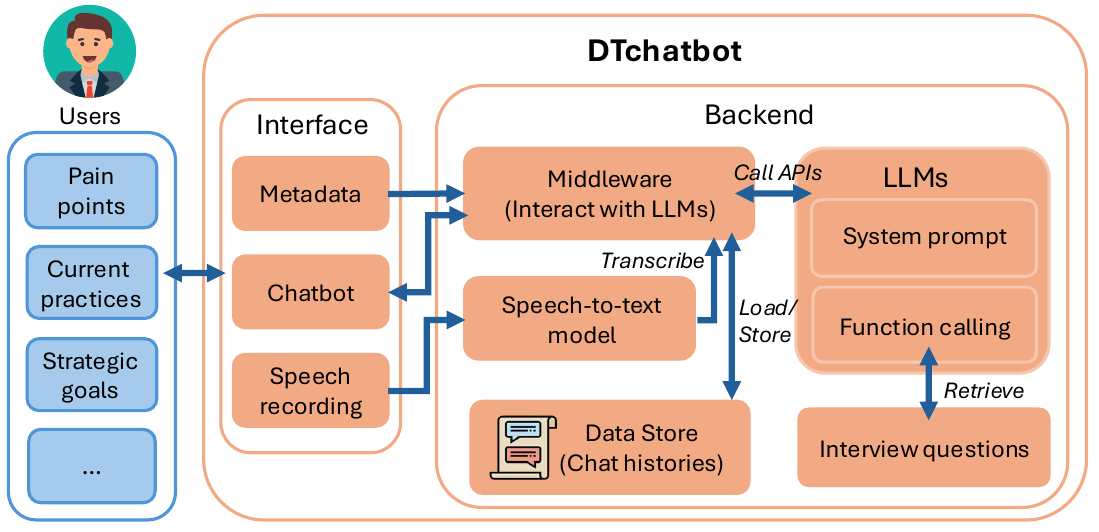}
    \caption{The architecture of the Consulting Chatbot.}
    \label{fig:archofBot}
\end{figure}

Our chatbot is fully developed based on open-source components, which allows easy adoption by other researchers and industries. The architecture of the DTchatbot is shown in Fig~\ref{fig:archofBot}, including two main components, \textit{chatbot interface} and \textit{backend}. Users, typically organisational representatives, such as product mangers or engineers, interact with DTchatbot through the \textit{interface} by answering questions that were designed to identify their digital transformation needs. We describe the details next.

\subsection{Chatbot interface}

It is designed with two primary features to facilitate the identification of digital transformation needs. First, it enables recording client information, such as company name, client name, industry type and size, and job title. This structured data allows for the categorisation of consulting workflows based on job titles, providing a tailored approach to addressing client needs. For instance, within a single organisation, expert consultations may involve professionals from various departments, and the job title serves as a critical attribute to distinguish the interactions. This categorisation ensures that the focus of the needs identification is aligned with the specific expertise of each client, fostering a more effective and targeted engagement. This recorded information also enables clients to revisit and continue conversations that were not completed in previous sessions, which is described in more detail in Section~\ref{sec:consulting bot}.  

Second, the interface supports multilingual and multimodal inputs, e.g., text and speech, ensuring accessibility and flexibility for diverse users. For textual interactions, the DTchatbot leverages the capabilities of LLMs, like OpenAI's GPT-4~\cite{ChatGPT}, which supports over 90 languages. This allows seamless communication across linguistic boundaries. Additionally, DTchatbot integrates an speech-to-text model, such as OpenAI Whisper model~\cite{openaiwhisper}, which supports over 50 languages~\cite{whisper}, to process voice inputs from users. This multimodal and multilingual support broadens accessibility and enhances the user experience (see Fig~\ref{fig:multilang}). This designed feature addresses the limitations of traditional workshops, particularly those stemming from language barriers and reliance on voice recording devices.  

\subsection{Backend}\label{sec:consulting bot}

\begin{figure}
    \centering
    \begin{subfigure}[b]{0.48\linewidth}
         \centering
         \includegraphics[width=\linewidth]{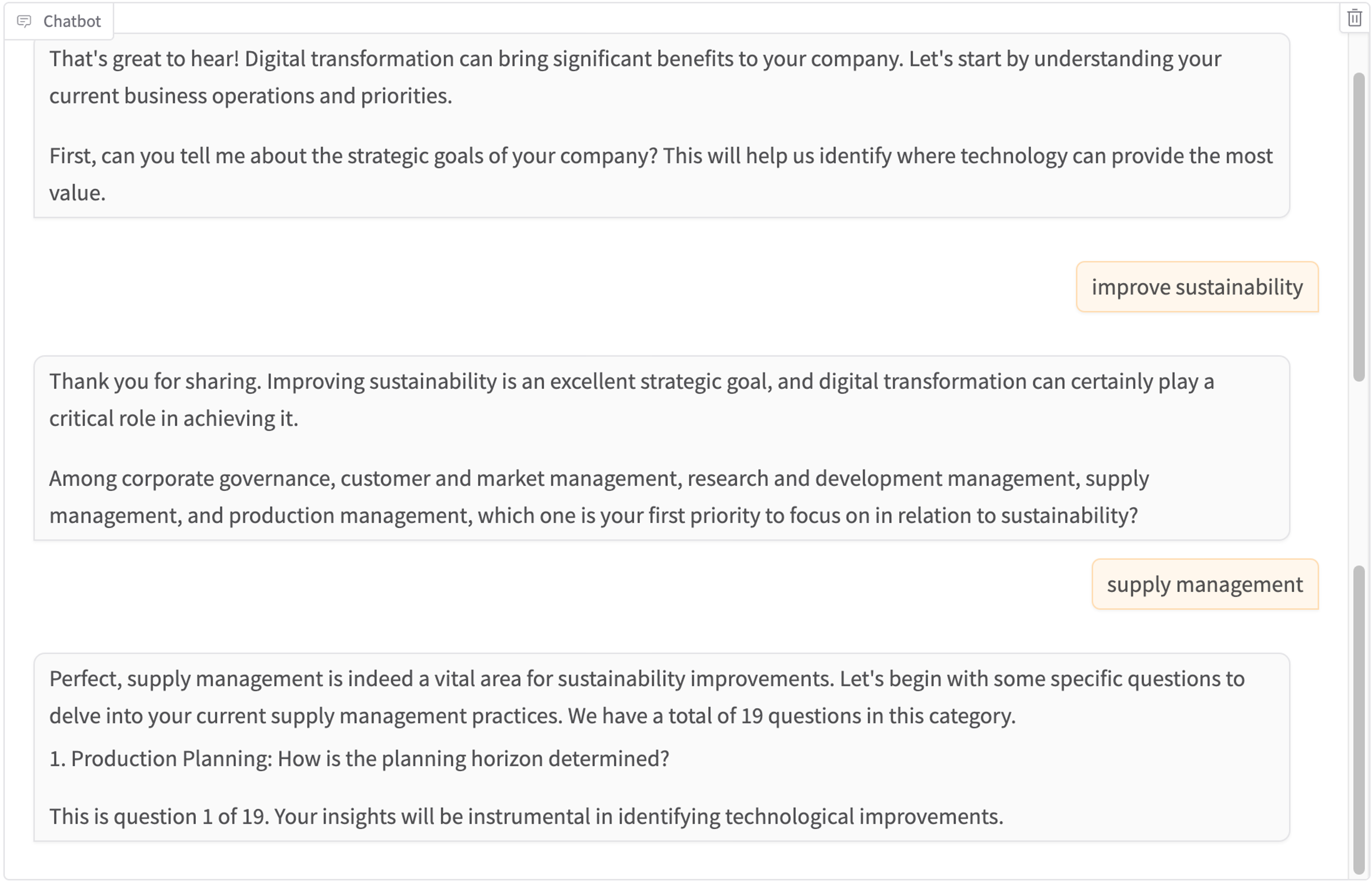}
         \caption{Demonstration of workflow-based interview questions.}
         \label{fig:demo1}
    \end{subfigure}
    \hfill
    \begin{subfigure}[b]{0.48\linewidth}
         \centering
         \includegraphics[width=\linewidth]{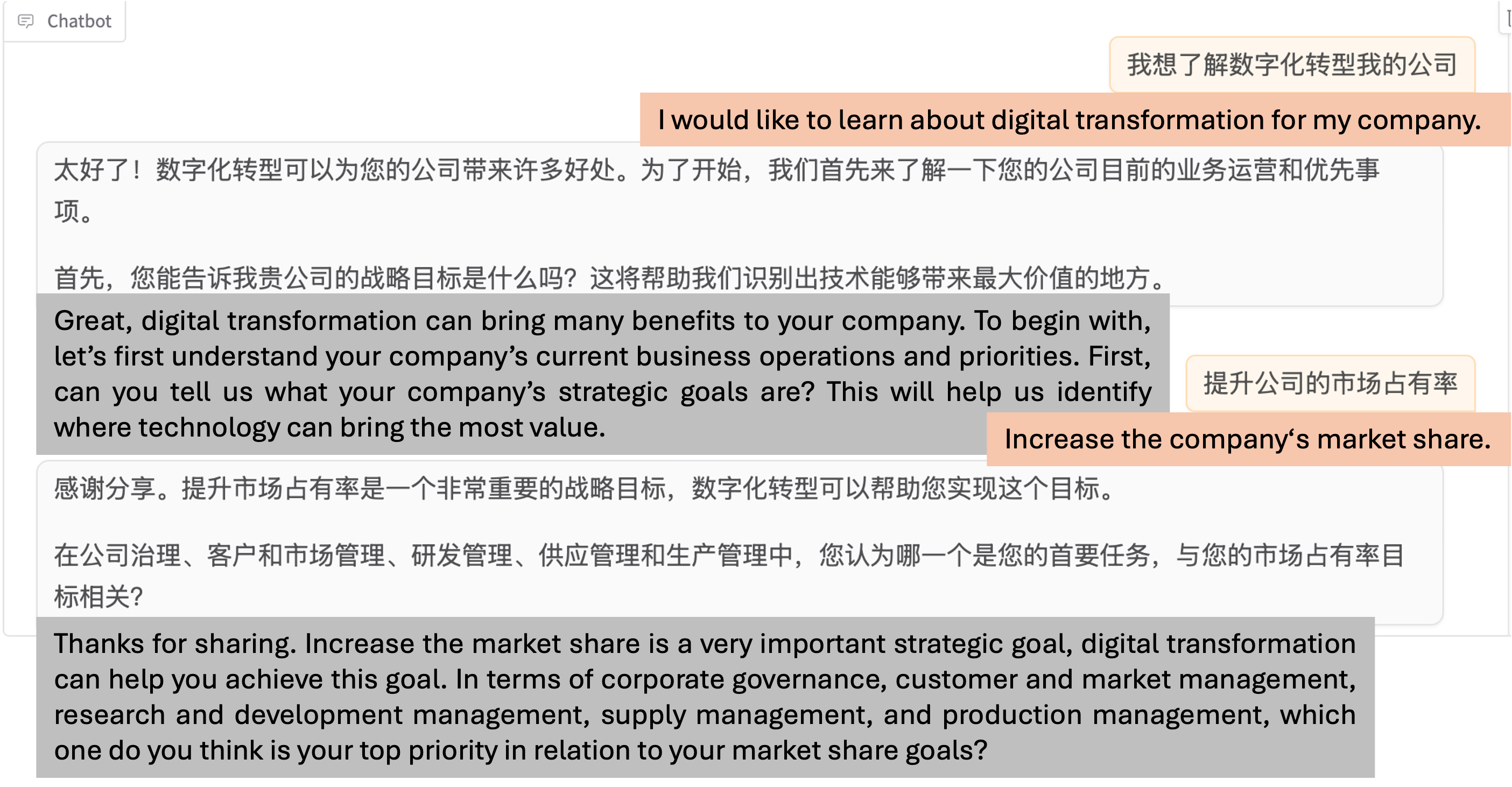}
        \caption{Demonstration of multi-language support (Chinese and its English translation).}
        \label{fig:multilang}
    \end{subfigure}
    \begin{subfigure}[b]{0.5\linewidth}
         \centering
         \includegraphics[width=\linewidth]{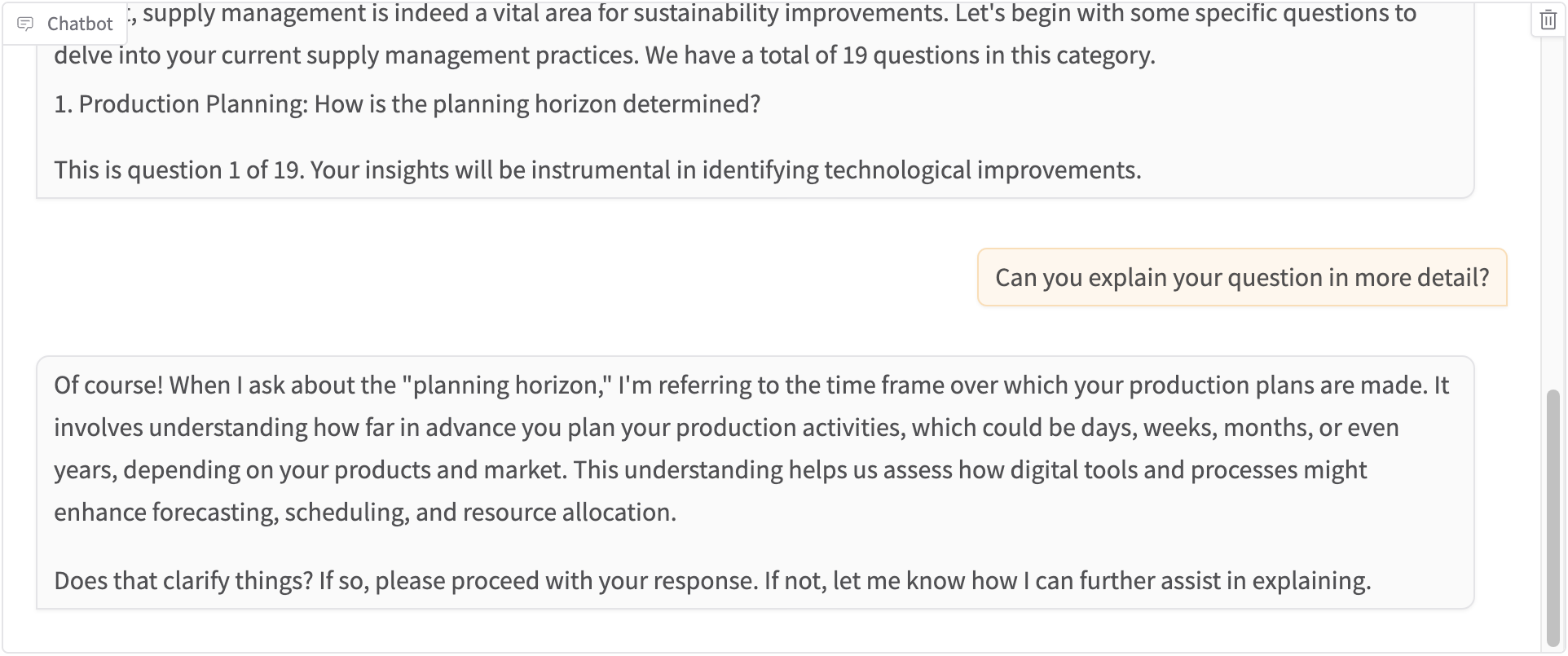}
        \caption{Demonstration of client guidance.}
        \label{fig:client guidance}
    \end{subfigure}
    \caption{Demonstration of DTchatbot}
    \label{fig:demo dtchatbot}
\end{figure}

The backend of the chatbot is instructed to ask questions based on a pre-defined workflow, ensuring a structured and systematic method for collecting information. The questions are designed and organised according to the workflow, which guides the conversation in a logical sequence. The workflow begins by asking users to identify their priorities across five key domains of digital transformation: \textit{corporate governance}, \textit{customer and market management}, \textit{research and development}, \textit{supply chain}, and \textit{production management}. Based on these priorities, the chatbot dynamically presents targeted interview questions within each domain to elicit organisation-specific information. Fig~\ref{fig:demo1} shows that the DTchatbot follows our predefined workflow to conduct interviews.

We adopt D3A questions, developed by Bogazici University, as our interview questions, which are used nationally by TUBITAK through certified consultants to evaluate firms’ digital transformation capability levels and create roadmaps~\cite{TUBITAK}. They are designed for in-depth interviews, with responses scored on a 0-4 scale. These questions form the core input for a chatbot aimed at streamlining digital transformation assessments. The chatbot uses the same logical flow to guide users, ensuring consistent and comprehensive data collection. The questions have been pilot-tested with SMEs working in Istanbul Dudullu and nearby industrial zones, including over 146 SMEs across key sectors (around 16,500 employees)~\cite{Bogazici}. The complete set of questions~\cite{Bogazici}, detailed for each category, is provided in the Appendix~\ref{sec:question list}.

By leveraging the capabilities of LLMs, DTchatbot provides real-time support for technical clarifications and guidance, improving the experience of the information collection process. For example, if a user mentions a need for implementing a cloud-based solution but is unsure about the differences between public, private, and hybrid cloud models, the bot can offer a clear explanation if the user asks a question like ``Can you explain to me the differences among public, private and hybrid cloud services?'' The chatbot can generate accurate responses. Similarly, if a client references a specific technology, without fully understanding its potential applications in a domain, the chatbot can provide tailored insights, examples, and even recommendations based on best practices. This enhances user engagement by making the experience more interactive, supportive, and personalized.

This capability is further expanded to provide guidance to users on how to effectively answer interview questions, ensuring that they articulate their needs and priorities. For instance, if a user is unsure how to respond to a strategic question, DTchatbot can offer examples or prompts to help them frame their objectives. By offering this guidance, DTchatbot not only improves the quality of the input but also encourages deeper reflection, enabling a more thorough and insightful exploration of their digital transformation needs. Fig~\ref{fig:client guidance} demonstrates this feature. 

Upon completing the conversation, DTchatbot automatically generates a comprehensive summary report outlining current practises, identified challenges, and strategic goals, providing a holistic view to inform digital strategy.

The full conversation history of the consulting process is stored on the web server in a structured format. The conversation history is systematically organised and labelled using key metadata, including the company name, interviewee name, and job title, as inputted through the user interface. This labelling ensures that each consultation session is uniquely identifiable and contextually relevant. By recording this information and storing the interaction history, users can seamlessly pick up where they left off, without the need to repeat or reintroduce their queries or responses. This feature not only enhances the user experience but also supports a more efficient and customised information collection, as the bot retains context and builds on previous discussions. The messages within the conversation are clearly distinguished, with labels explicitly indicating whether a message was provided by the user or generated by the DTchatbot. This structured format enhances traceability and simplifies subsequent analysis, such as tailored recommendations and iterative refinement of transformation strategies.

\subsection{Implementation}

We utilise Gradio~\cite{abid2019gradio}, an open-sourced framework for building web applications, to build the \textit{user interface}. Gradio provides a suite of off-the-shelf tools that accelerate the creation of responsive interfaces. Using Gradio, we design an interface that supports both text-based and audio-based interactions with the chatbot. This interface acts as the primary medium for collecting user input and delivering responses. 

The \textit{backend} includes a middleware that bridges the user interface and the LLMs. It processes and packages the text input into structured requests that are sent to the LLMs through APIs for processing. For audio inputs, we integrate the OpenAI Whisper model~\cite{openaiwhisper}, which transcribes spoken responses into text. The text is then sent to the middleware for communicating with LLMs. 

\begin{figure}[h]
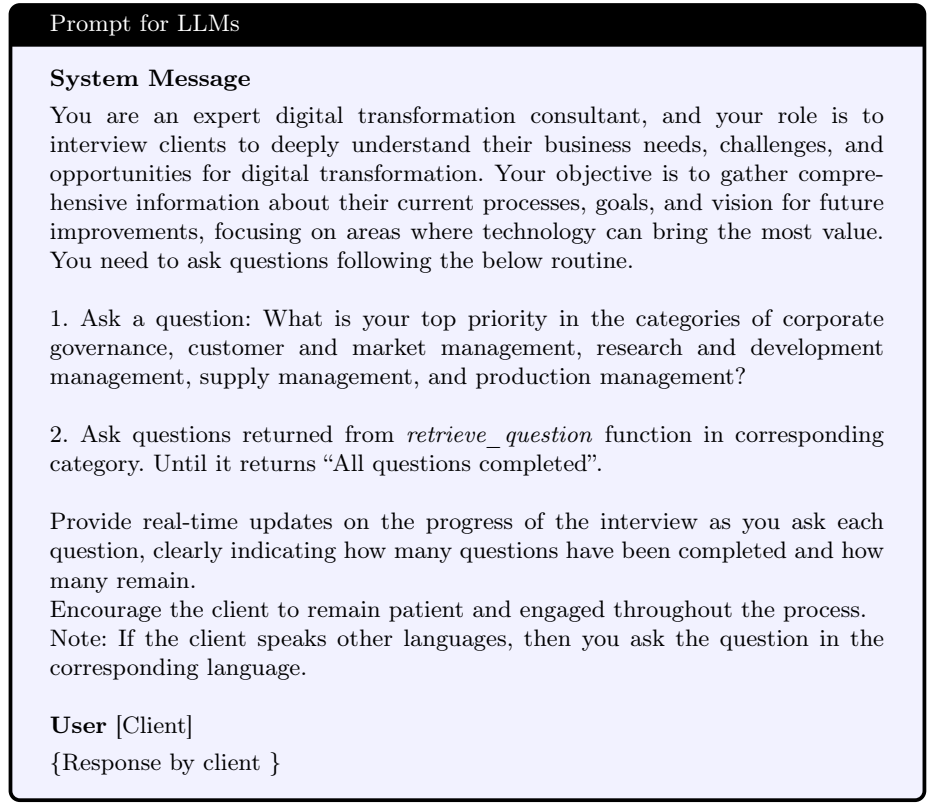

    
    \centering
    \begin{tcolorbox}
    [colback=blue!5!white,colframe=black,width=1.0\textwidth,title={Prompt for LLMs}]
    \small
    \textbf{System Message}
    \vspace{3pt} \\
    You are an expert digital transformation consultant, and your role is to interview clients to deeply understand their business needs, challenges, and opportunities for digital transformation. Your objective is to gather comprehensive information about their current processes, goals, and vision for future improvements, focusing on areas where technology can bring the most value. You need to ask questions following the below routine. \\

    1. Ask a question: What is your top priority in the categories of corporate governance, customer and market management, research and development management, supply management, and production management? \\

    2. Ask questions returned from \textit{retrieve\_question} function in corresponding category. Until it returns ``All questions completed''. \\
    
    Provide real-time updates on the progress of the interview as you ask each question, clearly indicating how many questions have been completed and how many remain. \\
    Encourage the client to remain patient and engaged throughout the process. \\
    Note: If the client speaks other languages, then you ask the question in the corresponding language. \\

    \textbf{User}\ [Client]
    \vspace{3pt} \\
    \{Response by client \}
    \end{tcolorbox}
    \caption{An excerpt of prompt for LLMs.}
    \label{tab:prompts for consulting bot}
\end{figure}

We use the capabilities of LLMs to conduct interviews aligned with predefined questions designed to acquire digital transformation needs. The bot operates based on carefully crafted system prompts, which guide its behaviour during the interview process. The function-calling capabilities of LLMs are utilised to dynamically retrieve interview questions from a predefined question list, ensuring seamless adaptability during the information collection process. This question list operates as a modular plug-in for the chatbot, allowing for effortless customisation and editing based on the specific stage of the consultation or the requirements of different interviews. An excerpt of the prompt is shown in Fig~\ref{tab:prompts for consulting bot}. We interact with LLMs via API request, enabling flexibility in selecting and utilising different LLMs as needed. Specifically, in our study, the OpenAI GPT-4 model is used to process user queries and generate responses. The chat scripts are stored on the Web server, referred to as the \textit{Data Store}, in a structured JSON format.

\section{Preliminary user study}

We conducted a preliminary user study to evaluate the capability of DTchatbot in identifying the digital transformation needs of organisations. The study involved two digital transformation experts and representatives from two SMEs, enabling us to gather insights from both expert and client perspectives. The primary objectives of the evaluation were to investigate both the effectiveness and the experience of using the chatbot following: \textbf{Obj1}: How do digital transformation experts perceive the use of a chatbot for acquiring organisations' needs? \textbf{Obj2}: How do users from SMEs reflect on their experience with the DTchatbot?

\textbf{Obj1} involves expert interactions with DTchatbot, where participants were asked to provide feedback (see Appendix~\ref{sec:expert opinions}) on two key aspects: 1) reduction of effort, and 2) enhancement of functionality and usability. For the \textbf{obj2}, the participants interacted with the DTchabot and they were asked to provide feedback, using the targeted questions (see Appendix~\ref{sec:clients opinions}). These questions focused on four key areas: \textit{perceived benefits}, \textit{interaction and usability}, \textit{data input and analysis}, and \textit{perceived limitations}. The profiles of SME representatives are: 1) human resource manager in an engineering solution company, 2) production director of a manufacturing company. Other information is detailed in Appendix~\ref{sec:profiles}.

\subsection{Expert's opinions}

Both experts agreed that the DTchatbot helps support digital transformation initiatives and effectively reduces effort. \textit{Expert 1} highlighted that the chatbot is user-friendly and offers a logical question sequence, which aids in the initial identification of gaps. While it is limited to detailed analysis, it provides a good starting point for understanding the specific needs of organisations. \textit{Expert 2} found the chatbot particularly useful for providing suggestions, strategies, and implementation steps, helping users focus on their transformation goals. The digital tool recommendations save significant time and effort by eliminating the need for extensive manual searches, enabling users to concentrate on higher-priority tasks.

Experts also suggested several enhancements to improve the chatbot’s functionality and usability:

\begin{itemize}
    \item \textbf{Interactive Features}: \textit{Expert 1} suggested re-organising the question sequence dynamically based on user inputs for better contextual alignment. \textit{Expert 2} recommended introducing selection boxes for common responses, reducing the need for manual input and making interactions more efficient.
    \item \textbf{Visualisation and Reporting}: \textit{Expert 2} proposed that the chatbot generate flowcharts to visually represent the suggested digital transformation processes, enhancing user comprehension. At the end of the conversation session, the chatbot could provide a detailed report summarising its recommendations to better assist users.
    \item \textbf{Analytical Depth}: Expert 1 suggested that integrating the analysis of diverse data types would enable the chatbot to deliver more detailed and insightful results. Moreover, including explanatory examples within the questions would improve user understanding and facilitate accurate responses.

\end{itemize}

\subsection{Client's reflections}

Client reflections are categorized into four areas as follows:

\noindent\textbf{Perceived benefits}: Participants found the chatbot interaction beneficial, particularly in raising awareness about their gaps in the early stages of digital transformation. They appreciated its ability to identify deficiencies and guide decision-making, highlighting its potential as a diagnostic tool that complements expert consultancy. One participant stated, ``It is beneficial for SMEs to raise awareness about understanding their gaps in the initial stages of the digital transformation. If there is an ongoing process, the chatbot might help observe areas for improvement''. 

\noindent\textbf{Interaction and usability}: All participants emphasised the voice transcription functionality and quick response time. These features were particularly appreciated for their ability to streamline interactions and reduce the effort required for communication. However, one participant suggested that introducing an \textbf{answer auto-completion} feature could further enhance the user experience. They stated, ``Suggestions could be provided for each question. For example, after writing short keywords for responses, automatic sentence completion or sentence improvement could enhance the interaction''. This feedback highlights the potential to integrate intelligent text assistance into the chatbot, allowing users to provide more structured and detailed responses with less effort.

\noindent\textbf{Data input and analysis}: The participants appreciated the ability of the DTChatbot to maintain meaningful conversations through logical follow-up questions. 
Additionally, they valued the chatbot’s capability to provide technical clarifications and guidance on answering questions, stating, ``It analyses customer complaints in-depth and provides answers, which is helpful''. However, participants also identified opportunities for improvement in usability. They suggested enhancing the chatbot by enabling it to accept diverse inputs beyond text, such as workflows or structured data entries. One participant remarked, ``Instead of answering the chatbot’s questions with text, allowing inputs such as workflows and data entries would make it more user-friendly and practical for generating digital solution suggestions''.

\noindent\textbf{Perceived limitations}: 
The participants highlighted that although the chatbot is effective for initial awareness and identifying gaps, it may not be sufficient to fully support digital transformation efforts. They stated, ``It is good for understanding ourselves in the initial stages, but it is not sufficient for a complete digital transformation process. For example, it could be integrated into a digital system like an Enterprise Resource Planning (ERP) to increase usability''.

It is important to note that data privacy and security are critical considerations for the DTChatbot, given the sensitive nature of organisations' current practices and challenges discussed during consultations. To address these concerns, the DTChatbot ensures that all client data, including chat histories, is securely stored with a robust access control system.

\section{Discussion and Future Work}

The DTchatbot offers significant advancements in the process of eliciting digital transformation needs through its multimodal and multilingual capabilities. It provides a highly scalable and accessible solution for acquiring information. Its integration of advanced LLMs, ensures context-aware guidance and technical clarification, providing a highly personalised and insightful consulting experience. Additionally, its structured workflow-based design ensures that consultations are systematic and aligned with pre-defined instructions. The ability to record interaction histories facilitates downstream tasks, like the generation of actionable insights and detailed reports. The integration of the OpenAI Whisper audio-to-text model streamlines the process by enabling seamless handling of audio inputs, enhancing overall efficiency.

This work demonstrates the potential of LLM-powered chatbots to provide dynamic, context-aware, and domain-specific interactions. Beyond digital transformation, the structured and interactive capabilities of the DTChatbot suggest broader applicability to general expert interviews and consulting tasks across industries, underscoring its versatility as a tool for knowledge acquisition and user engagement. It highlights the implications of employing chatbots to address the challenges of engagement and scalability.

Despite these advancements, several challenges and constraints remain. One critical concern is ensuring data privacy and security, as the DTchatbot processes sensitive client information. Addressing this limitation requires the implementation of strict data governance practices, including the deployment of locally hosted LLMs to mitigate risks associated with data disclosure. Moreover, while LLMs are powerful, they can occasionally generate responses that are inaccurate or overly generic, particularly when faced with highly specialized or nuanced queries. This highlights the need for fine-tuning LLMs to better cater to specific domains, such as digital transformation.

Future work will focus on accommodating reflections from both experts and SME users, such as integrating multiple data input formats, like images, documents, etc., to further improve usability. We also plan to extend its functionality to support detailed analysis, enabling the generation of tailored reports for organisations.

\bibliographystyle{splncs04}
\bibliography{sample-base}

\newpage
\appendix

\section*{Appendix}
\section{Interview question list}\label{sec:question list}

\begin{longtable}[c]{| p{.25\textwidth} | p{.75\textwidth} |}
\caption{The list of interview questions in each category.} 
\label{tab:allQuestions} \\
\hline
Category   & Question\\ \hline
\endfirsthead
\hline
Category   & Question\\ \hline
\endhead

\hline
\endfoot

\hline
\endlastfoot

\multirow{12}{=}{\raggedright Corporate Governance} 
    & How are management decisions made? \\
    & Is there a written strategic plan? \\
    & Is there a strategy for digitalisation? \\
    & Are business processes defined? \\
    & What is the level of cooperation between units? \\
    & How are financial records created? \\
    & Who is responsible for information systems? \\
    & Are there cybersecurity systems in the organisation? \\
    & Is it possible to access company data remotely? \\
    & How is the training and development of employees managed? \\
    & What is being done to improve digital competencies of employees? \\
    & How are new ideas collected within the company? \\
\hline

\multirow{12}{=}{\raggedright {Customer and Market Management}} 
    & How are sales and marketing activities carried out? \\
    & How do you make sales forecasts? \\
    & How is sales data shared with other business units? \\
    & How do you create quotes? \\
    & What can your customers do through digital media? \\
    & How are customer conversations and related information stored? \\
    & How do you receive orders from your customers? \\
    & How is the dealer network monitored? \\
    & How are customer projects tracked? \\
    & How is the sales team's performance monitored? \\
    & How is distributor performance monitored? \\
    & How do you track customer feedback and issues such as after-sales returns, technical service, complaints? \\
\hline

\multirow{10}{=}{\raggedright {Research and Development Management}} 
    & Is there a P\&D or R\&D department? \\
    & Are there any patents and/or patent applications? \\
    & Is there any cooperation with academic institutions in product development and innovation? \\
    & Is there an externally supported R\&D project? \\
    & How are product/process/material design - development   - engineering studies done?  \\
    & Who is involved in product/process/material design -   development - engineering studies? \\
    & Do you produce the technologies used or buy them ready-made? \\
    & Is there hardware on products to collect data (e.g., sensors, chips)? \\
    & ow is the need to develop new products, new services   and processes determined? \\
    & Is it possible to customise the product? \\
\hline

\multirow{19}{=}{\raggedright {Supply Management}} 
    & Production Planning: How is the planning horizon determined? \\
    & Production Planning: How is production planning done? \\
    & What happens when it is necessary to make changes in   the production plan due to a disruption or need caused by the supplier, customer or institution?  \\
    & How are batch sizes determined in production? \\
    & How is capacity management carried out in the enterprise? \\
    & How do you determine material requirements? \\
    & Purchase Orders: How do you purchase materials? \\
    & Purchase Orders: How is communication with suppliers ensured? \\
    & How do you choose suppliers? \\
    & How is supplier performance evaluated? \\
    & How is information shared with other company functions (sales, purchasing, production, storage, shipment)? \\
    & How is information shared with external supply chain partners (suppliers, logistics, customers)? \\
    & How is raw material stock planning done? \\
    & How are raw material stocks tracked? \\
    & How is warehouse management done (raw materials, components, finished products)? \\
    & How is line feeding done: physically conveying materials? \\
    & How is line feeding done: material demand of   production? \\
    & How are logistics work orders created? \\
    & How are logistics managed? \\
\hline

\multirow{20}{=}{\raggedright {Production Management}} 
    & How do you forward production work orders to the line? \\
    & How do you forward bills of materials to production? \\
    & How is scheduling/rescheduling done? \\
    & How do you track production? \\
    &  How do you monitor machines and downtime during production? \\
    & How is workforce monitoring done? \\
    & How do you keep track of material movements in production area? \\
    & How do you monitor production performance? \\
    & How is production information shared internally? \\
    & How is Quality Management done? \\
    & How do you handle quality problems related to materials, products and/or processes (problem assessment)?  \\
    & How do you handle quality problems related to materials, products and/or processes (nonconformity record)?  \\
    & How do you evaluate quality control data related to your materials? \\
    & How do you evaluate quality control data related to your finished products? \\
    & How do you evaluate quality control data related to your processes? \\
    & How do you evaluate quality control data related to your semi-finished products? \\
    & How is maintenance management carried out? \\
    & What methods are used for machine maintenance? \\
    & How are maintenance planning and scheduling done? \\
    & How is energy consumption monitored? \\
\hline   

\end{longtable}

\section{Questions for collecting experts' opinions}\label{sec:expert opinions}

\begin{enumerate}
    \item How effective is the DTchatbot in reducing user effort and supporting digital transformation?
    \item What additional features and functionalities could improve the usability and analytical capabilities of the DTchatbot?
\end{enumerate}

\section{Questions for collecting SME clients' opinions on DTchatbot}\label{sec:clients opinions}
\begin{enumerate}
    \item Perceived Benefits
    \begin{itemize}
        \item Did the chatbot help you identify gaps in your digital transformation journey? How?
        \item Were the chatbot’s questions helpful in understanding your organisation’s needs?
    \end{itemize}
    \item Interaction and Usability
    \begin{itemize}
        \item Was the chatbot easy to use? What aspects could be improved?
        \item Did the follow-up questions enhance the conversation?
    \end{itemize}
    \item Data Input and Analysis
    \begin{itemize}
        \item Would it be useful if the chatbot allowed different types of inputs, like workflows or data files?
        \item How should the chatbot process and present information to better meet your needs?
    \end{itemize}
    \item Perceived Limitations
    \begin{itemize}
        \item Is the chatbot sufficient for a complete digital transformation, or does it need additional features (e.g., system integration)?
    \end{itemize}
    \item General Feedback
    \begin{itemize}
        \item What single improvement would make the chatbot more useful for your organisation?
    \end{itemize}

\end{enumerate}

\section{Participants Profile}\label{sec:profiles}

Digital transformation experts:
\begin{itemize}
    \item Expert 1: Early career researcher and academic in Supply Chain Management, expert on Industry 4.0, digital transformation, sustainability, sustainable supply chain management.
    \item Expert 2: Senior researcher and academic in Design and Innovation, expert on Design creativity, product innovation, design management.
\end{itemize}

\noindent Representatives of SMEs:

\begin{itemize}
    \item The Human Resource Manager of an SMEs specialising in advanced engineering solutions for the defence industry, including remote-controlled detonators and unmanned systems.
    \item The Production Director of a manufacturing SME known for its production of engine cooling systems, including viscous fan clutches and system solutions for industrial and automotive applications.
\end{itemize}

\end{document}